\documentclass{aa}

\usepackage{epsfig,latexsym,longtable,lscape}

\begin{document}

   \thesaurus{3
                 (11.19.2;  
                  11.19.5;  
                  11.19.7)}  

\title{Growth of Galactic Bulges by Mergers: I Dense Satellites}
\author{J. A. L. Aguerri$^{1,2}$, M. Balcells$^{1}$ \& R. F. Peletier$^{3}$}
\offprints{J.A.L. Aguerri}
\institute{1.-Instituto de Astrof\'{\i}sica de Canarias, E-38200 La Laguna, Tenerife, Canary Islands, Spain.\\
2.-Astronomisches Institut der Universitat Basel, CH-4102 Binningen, Switzerland.\\
3.-School of Physics and Astronomy, University of Nottingham, NG7 2RD, UK\\
email: jalfonso@astro.unibas.ch, balcells@ll.iac.es, Reynier.Peletier@nottingham.ac.uk }

   \date{Received ; accepted }
  \authorrunning{Aguerri, Balcells \& Peletier}
   \titlerunning{Galactic bulges and mergers}
   \maketitle

\begin{abstract}
Andredakis, Peletier \& Balcells (1995) fit Sersic's law 
$\mu(r) \sim r^{1/n}$ to the bulges of the Balcells \& Peletier 
(1994) galaxy sample, and infer that $n$ drops with morphological type T 
from $n \approx$ 4--6 for S0 to $n=1$ (exponential) for 
Sc's.   We use collisionless N body simulations to test the assumption
that initially the surface brightness profiles of all bulges 
were exponential, and that the steepening of the profiles toward the 
early-types is due to satellite accretion.   
The results are positive.  After the accretion of a satellite, bulge-disk fits
show that the bulge grows and that the bulge profile index $n$ increases 
proportional to the satellite mass.  For a satellite as massive as the bulge,
$n$ rises from 1 to 4.   We present kinematic diagnostics on the remnants and disk thickening. 
The latter suggests that the bulge growth must have occurred before 
the last formation of a thin disk in the galaxy.  The thick disks created by the merger are reminiscent of thick disks seen in early-type edge-on galaxies. 
 The efficiency of the process suggests that present day bulges of late-type spirals showing exponential profiles cannot have grown significantly by collisionless mergers.
\end{abstract}


\section{Introduction}
\label{Sec:Introduction}

Two paradigms are commonly used to explain the formation of the central 
bulges of disk galaxies.  In the first, the bulge forms prior to the disk, 
perhaps as part or an aftermath of the formation of the stellar halo 
(Gilmore \& Wyse 1998), or as a result of early merging resulting in an 
$r^{1/4}$ spheroid around which a new thin disk forms from surrounding gas 
(eg. Kauffmann, Guiderdoni \& White 1994).   In the second, bulges  form 
and grow after bar instabilities draw disk stars and gas to the center 
(Pfenniger \& Norman 1990).  Details about these paradigms are given in 
the review by Wyse, Gilmore \& Franx (1997). Bouwens, Cay\'on \& Silk 
(1999) perform a comparative test of these hypotheses. 

What is the role of interactions, mergers and accretion of satellites in 
the functioning of these processes?.  Satellite accretion must have 
occurred several times in a disk galaxy over a Hubble time.  While mergers 
of galaxies with similar mass fully destroy the disk (eg. Barnes \& 
Hernquist 1991) and leave remnants resembling an elliptical galaxy, 
satellite accretion does less damage to the disk;  and, if the satellite 
is dense enough, it must reach the galaxy center. The accretion in this 
case may drive the growth of the central bulge.  Several studies have 
addressed whether mergers are responsible for pushing a spiral galaxy 
along the Hubble sequence toward Sa/S0 (eg. Walker et al 1996).  Mergers 
must have been efficient in groups and in the field; in the cluster 
environment, disk cosmological fading due to the evaporation of neutral 
gas,  the cluster tidal field (Dubinski 1998) and occasional galaxy-galaxy 
encounters (Moore et al. 1996) contribute as well to the growing 
prominence of the bulge.

Satellite accretion onto disk galaxies has been extensively studied in the 
past using N body techniques.  The focus of these studies have been the 
thickening of the disk (Quinn \& Goodman 1986, T\'oth \& Ostriker 1992, 
Quinn, Hernquist \& Fullagar 1993, hereafter QHF93, Walker et al 1996, 
Huang \& Carlberg 1997), the formation of counterrotating disks in spirals 
(Thakar \& Ryden 1996), the formation of X structures in S0 galaxies 
(Mihos et al. 1995), and the triggering of nuclear starbursts (Mihos \& 
Hernquist 1994).  Diagnostics related to the central bulge are largely 
ignored.  Indeed, except for the latter paper, none of the studies 
mentioned above include a central bulge in their initial galaxy model, a 
bold simplification given that the potential of the central bulge is a key 
ingredient of the merger dynamics of both stars (Hernquist 1993) and gas 
(Mihos \& Hernquist 1994).  

Sufficient data exists nowadays on bulges (Balcells \& Peletier 1994, 
hereafter BP94, Peletier \& Balcells 1996, 1997, Peletier et al. 1999, 
Carollo 1999, de Jong 1996, Prieto et al. 1999, Ratnatunga, Griffiths \& 
Ostrander 1998, Schade et al. 1996, Abraham et al. 1999, Marleau \& Simard 
1998) to allow the structure and dynamics of bulges to be used as 
constraints on the accretion history of disk galaxies.  However, no 
detailed studies of the growth of bulges by accretion of satellites have 
been published to date.  

 In this paper we study whether the shape of the surface brightness 
profiles of bulges changes with the merger history of 
spiral galaxies.  Andredakis, Peletier \& Balcells (1995, hereafter APB95) 
show that bulges follow a systematic behavior in the $n$ vs. T plane, 
where $n$ is the exponent index in Sersic's law $\mu(r) \sim r^{1/n}$, and 
T is the galaxy type index given eg. in the RC3.  APB95 find that $n$ is 
4--6 in early, S0 types, decreasing to $n \approx 1$ for late types (Sc). 

The APB95 result could result from the effect of the disk potential on a 
bulge initially following an  $r^{1/4}$ law, in an scenario in which the 
bulge forms prior to the disk.  Andredakis (1998) studies the adiabatic 
growth of a disk onto an existing $r^{1/4}$ spheroid, and finds that the 
disk potential does modify the bulge surface brightness profile, lowering 
the $n$ in the exponent of Sersic's law.  But the mechanism saturates at 
$n=2$, showing that exponential bulges ($n = 1$) cannot be produced by 
adiabatic growth of the disk around an $r^{1/4}$ bulge. 

That merging is an efficient way to generate $r^{1/4}$ surface
brightness profiles is well-known (Gerhard 1981; Barnes 1988).  We may
thus conjecture that the behavior found by APB95 reflects the effects
of the relaxation in an scenario of bulge growth by accretion.  
 Here we address the following question: assuming that bulges are formed with exponential profiles, does the accretion of dense satellites simultaneously drive the growth of the bulge mass and the evolution of the bulge surface density profile toward higher-$n$ Sersic law shapes? 
In cosmological models of galaxy formation involving gas and stars,
central bulges show characteristic exponential surface density profiles
(Dom\'\i nguez-Tenreiro et al. 1999), suggesting that bulges may start
out with exponential profiles, and that any changes are imprints of
subsequent evolution.  To test the bulge growth conjecture we consider
disk-bulge-halo galaxy models for which the bulge surface brightness
profile is initially exponential, and fit Sersic's law to the bulge
surface brightness profile after an accretion of a satellite.  We then
draw "growth vectors" in the $n$ vs. T plane.

Our results suggest that accretion of dense satellites is efficient in 
turning exponential bulges into $r^{1/n>1}$ bulges.  After the merger, the 
$n$ index of Sersic's law increases proportional to the satellite mass.  
It reaches $n=4$ for a satellite as massive as the original bulge.  Our 
models suggest that this result is largely independent of the internal 
structure of the satellite, as long as the satellite is dense enough to 
reach the center undisrupted. We quantity disk thickening.  We concur with 
other studies of spiral galaxy evolution via merging, that only if the 
thin disk is rebuilt after the accretion can we match real galaxies.  We 
analyze the kinematic structure of the merger remnants to provide 
additional tests of the model.  For direct mergers, the rotation curve 
shapes are steeper for more massive bulges, consistent with observations.  
Massive satellites on retrograde orbits result in counterrotating bulges, 
suggesting either that massive accretion events are rare, or that 
accretion preferentially occurs on prograde orbits.  

\begin{table*}
\begin{center}
\caption{\label{Tab:ParsMain} Initial parameters of the main galaxy}
\begin{tabular}{ccccccccccccc}
\hline
\hline
\multicolumn {3}{c} {Bulge} & & \multicolumn {5}{c} {Disk} & & \multicolumn {3}{c} {Halo}  \\
\cline{1-3}\cline{5-9}\cline{11-13}
$M_{B}$ & $r_{B}$ & $N_B$ & & $M_{D}$ & $h_{D}$ & $R_{D}$ & $z_{D}$ & $N_D$ & &$M_{H}$ & $R_{H}$ & $N_H$ \\
(1) & (2) & (3) & & (4) & (5) & (6) & (7) & (8)& &(9) & (10) & (11) \\
\hline
0.422 & 0.24 & $10^{4}$ & & 0.82 & 1.0 & 5.0 & 0.10 & $4 \times 10^{4}$ & & 5.14 & 21.81 & $5 \times 10^{4}$ \\
\hline
\end{tabular}
\end{center}
\tiny
(1) Bulge mass
(2) Bulge half-mass radius (3) Number of bulge particles 
(4) Disk mass (5) Disk scale lenght (6) Disk truncation radius (7) Disk scale height (8) Number of disk particles (9) Halo mass (10) Halo truncation radius (11) Number of Halo particles.
\end{table*}

Model details are given in \S~\ref{Sec:Models}.  The results on the shape 
of the bulge surface brightness profiles are given in 
\S~\ref{Sec:Results}.  
Model kinematics are presented in \S~\ref{Sec:Kinematics}.
Section~\ref{Sec:DiskHeating} briefly presents 
results on disk heating.  Implications are discussed in 
\S~\ref{Sec:Discussion}.  A summary of results is given in \S~\ref{Sec:Summary}.  Throughout the paper, $M$ denotes mass and not absolute magnitude.  

\begin{figure*}[!htb]
    \begin{flushleft}
     \caption[Fig:Snap1]{(See aguerri.fig1.gif file) The time evolution of the luminous matter for model 1. Views from the $z$, $x$ and $y$ axes are given. At $t=0$, the satellite is at $(x,y,z)=(8.66, 0., 5.)$.  Orbit is counterclockwise, prograde with the disk rotation.} 
\label{Fig:Snap1}
    \end{flushleft}
    \end{figure*}

\begin{figure*}[!htb]
    \begin{flushleft}
    \caption[Fig:Snap5]{(See aguerri.fig2.gif file) The time evolution of the luminous matter for model 5. Views from the $z$, $x$ and $y$ axes are given. At $t=0$, the satellite is at $(x,y,z)=(8.66, 0., 5.)$.  Orbit is counterclockwise, prograde with the disk rotation. \label{Fig:Snap5}}
    \end{flushleft}
    \end{figure*}

\section{Models}
\label{Sec:Models}

The primary galaxy is modeled with the 
bulge-disk-halo model of Kuijken \& Dubinski (1995, hereafter KD95). 
The bulge is modeled as a King model.  This model was chosen for the bulge 
because its surface density profile is exponential between the core radius 
and the truncation radius, providing a good match to the surface 
brightness profiles of late-type spirals, and because it is spheroidal.  
The core radius is 0.15 (units given below), and the concentration 
parameter, taken as the ratio of the bulge truncation radius and the core 
radius, is 6.7.  The disk surface density is exponential both in the 
galactic plane and in the perpendicular direction to the plane. The scale 
height is ten times less than the scale length (see eg. Guthrie 1992, de 
Grijs 1998).  The chosen disk velocity dispersion makes the disk warm with 
a Toomre $Q=1.7$ at the disk half mass radius.  The value of $Q$ is fairly 
constant throughout the disk, though rising both in the center within 1 
scale length and near the edge (KD95). This $Q$ parameter does not allow 
the growing of bar-type perturbations in the disk when in isolation. The 
halo has a distribution function of an Evans model (Kuijken \& Dubinski 
1994). 

The model used for our experiment is equal to model $A$ of KD95  This 
model matches the Milky Way when the units of length, velocity and mass 
are $R= 4.5$ kpc, $V=220$ km s$^{-1}$, $M=5.1 \times 10^{10} M_{\odot}$. 
Note that the core radius is then 0.71 kpc, indicating that the models 
lack central resolution.   Masses, radii, and number of particles for each 
component are given in Table~\ref{Tab:ParsMain}. A gravitational constant 
of $G=1$ is used throughout. 

The satellite galaxy is modeled either as a non-rotating Hernquist sphere 
(Hernquist 1990), with outer radius similar to that of the bulge, or with 
a King model.  The King model satellite is useful to verify that the 
accretion-driven evolution toward an $r^{1/4}$ profile is not motivated by 
the fact that the 
Hernquist satellite models already have a surface density profile 
approaching the $r^{1/4}$ law. 

\begin{figure*}[!htb]
    \begin{flushleft}
\psfig{figure=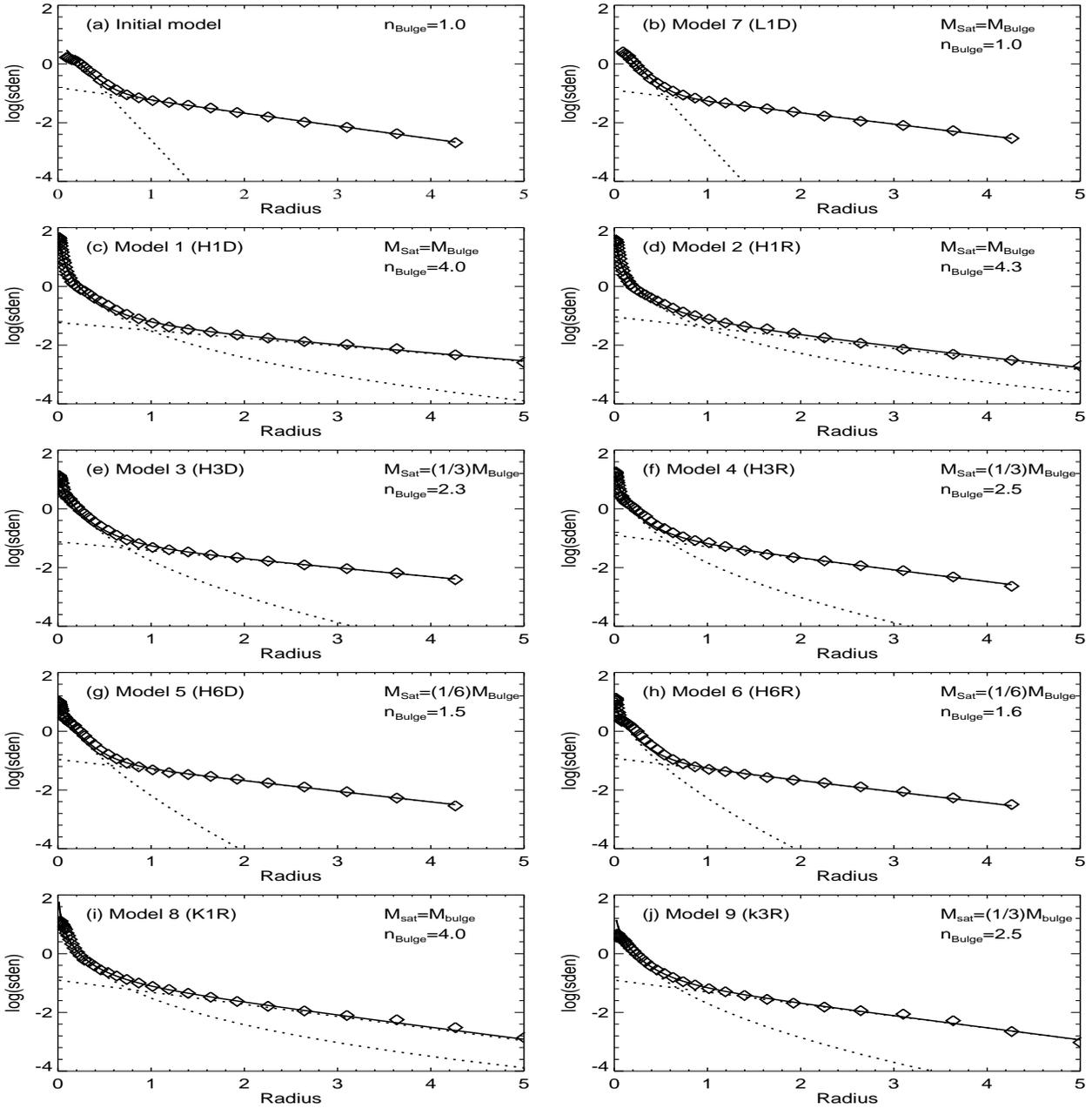,height=17.truecm,width=17.truecm}
              \caption[Fig:SBfits]{Radial surface density distribution of luminous matter.  Error bars are smaller than the symbol size.  
{\it (a)}  Initial model. 
{\it (b)} Low-density model 7 after the merger is complete.  
{\it (c--h)} Models 1--6 after the merger is complete.  
{\it Points: } model measurements.  
{\it Dotted lines: } Exponential and $r^{1/n}$ fitted components. 
{\it Solid line: } Sum of the two fitted components.  \label{Fig:SBfits}}
    \end{flushleft}
    \end{figure*}

Both the disk and the satellite galaxy models are allowed to relax
separately for 1.5 disk rotation periods (measured at 1 disk scale length)
before starting the simulations.  The surface density profile of the
initial disk galaxy model after relaxation is shown in
figure~\ref{Fig:SBfits}a.  Its projected rotation curve is shown in
figure~\ref{Fig:Rotcur}a together with the model's circular velocity curve. 
We ran the disk galaxy model in isolation for 2000 time updates (14 disk
rotation periods at 1 disk scale length, the maximum duration of our merger
experiments).  The shape of the bulge surface density profile is stable
(same $n$ value at start and at end of run), while the disk thickens in
close agreement with the thickening results of KD95.  This demonstrates
that, while the number of particles is admittedly low, the results do not
suffer from two-body relaxation as the more massive halo particles cross
the disk and the bulge.  This result also demonstrates the excellent
stability properties of the KD95 model, which makes it ideally suited to
experiments of interaction-driven structural evolution such as the ones in
this paper.

We run several merger experiments varying the mass of the satellite 
galaxy.  We explored three values of the bulge-to-satellite mass ratio, 
1:1, 3:1, and 6:1.  For each satellite mass setting we run a direct orbit 
(inclination 30$^\circ$ w.r.t. the disk plane) and a retrograde orbit 
(inclination 150$^\circ$).
Initial orbits were elliptical with apocenter equal to twice the disk 
outer radius and pericenter equal to twice the diskscale length.  
 We don't explore the dependency of the results on the orbital
energy or angular momentum.  This choice is probably restrictive. 
We expect a dependency on orbital initial conditions to be significant for the
disk (eg. QHF93), though possibly not for the bulge; when the satellite
reaches the inner parts of the galaxy, the potential's circular
velocity should determine the energy of the bulge-satellite merger more so
than the orbit initial conditions. 
For all these models a Hernquist satellite was used.  Retrograde merger orbits 
with mass ratios 1:1 and 1:3 were repeated using a King model satellite.  
While full exploration of satellite density effects is beyond the scope of 
this paper, we run one model in which the satellite has low density (model 
7 in Table~\ref{Tab:ParsOrbit}).  This model is similar to the models of 
Velazquez \& White (1999).  Orbital parameters for the merger experiments 
are shown in Table~\ref{Tab:ParsOrbit}.  Satellite masses and half-mass 
radii are given in Table~\ref{Tab:ParsOrbit}. We used a $M \sim r^{1.3}$ 
scaling between satellites of different masses.  In the remainder of the 
paper, models are referred to using a three-character code: {\tt SMO}, 
where {\tt S} describes the satellite (H for Hernquist, K for King, L for 
low-density), {\tt M} describes the ratio of bulge mass to satellite mass 
(values 1, 3, 6), and {\tt O} describes the orbit (D for direct, R for 
retrograde).  Codes are given in Table~\ref{Tab:ParsOrbit}.  On one of the merger remnants, we ran a second merger, which we denote H3R3R (see \S~\ref{Sec:Results} and inset in Figure~\ref{Fig:NvsMrat}).  We also
attempted a multi-merger experiment involving 10 small satellites 
(\S~\ref{Sec:Discussion}).  

Computations were carried out with a SGI Power Challenge machine 
(6 64-bit R8000 processors).  Evolution was computed using the TREECODE of 
C. H. Heller (see Heller 1991, Heller \& Sholsman 1994), kindly made 
available by the author.  Heller's code, with SPH turned off, 
uses the cubic tree structure 
described by Barnes \& Hut (1986). The algorithm updates the particle 
positions with the leap-frog algorithm, with a variable time step ranging 
between 0.01 and 0.05. The gravitational force was softened with a spline 
kernel (Hernquist \& Katz 1989) with constant softening length $\epsilon = 
0.02$.  No quadrupole-moment corrections were applied. In cubic treecodes, 
a monopole-only calculation increases force errors by about 50\% relative 
to a calculation including quadrupole terms (Hernquist 1987).  Hence, for 
the tolerance parameter used ($\theta=0.8$), our code computes forces 
within 1.5\% of those given by direct summation, compared to typical 1\% 
errors of the calculation to quadrupole order.   All models were evolved 
beyond the full merger.  Total energy was conserved to better than 0.1\% 
during the simulation.   Two of the merger models were repeated using 
Hernquist's version of the TREECODE (Hernquist 1987, 1990a).  The final 
surface density profiles were identical to those computed with Heller's 
code.  

\section{Results}
\label{Sec:Results}

In figures \ref{Fig:Snap1} and \ref{Fig:Snap5} we show snapshots of the 
evolution of the models H1D and H6D.  In both these models, 
as in models 1--6, 8 and 9, the core of the satellite reaches the center of the 
primary galaxy and merges with the bulge.  The scaling of dynamical 
friction with the satellite mass causes the merger times to significantly 
increase as the satellite mass decreases.  The merger affects the 
structure of the main galaxy in several ways.  Transient spiral patterns, 
warps and non-axisymmetrical structures appear in the disk of the main 
galaxy.  We address those in \S~\ref{Sec:DiskHeating}.  

We first focus on the structural parameters of the bulge. 
Figure~\ref{Fig:SBfits}a shows the surface density profile of the main
galaxy before the merger.  Poisson-based error bars in the density, not plotted, are a few percent at most owing to the large number of particles.  
We decompose this profile in two components much
in the same way as observers decompose a galaxy surface brightness
profiles.  The disk portion of the profile is well fit with an exponential
law.  The Sersic law written as

\begin{equation}
I(r) = I_e \exp \left[ -(0.868n - 0.142) \left\{  \left( \begin{array}{c} 
\frac{r}{r_e} \end{array} \right)^{1/n} - 1 \right\} \right]
\end{equation}

is used for the bulge (S\'ersic 1968; Andredakis et al 1995; Prieto et al. 
1999), with $I_e$, $r_e$ and $n$ as free parameters.  The best fit is
achieved with $n=1$, i.e. exponential profile.  Deviations occur only at
radii smaller than the core radius, where the surface density profile
flattens.

Figures~\ref{Fig:SBfits}b--j show the final face-on,
azymuthally-averaged radial surface density profiles of all luminous matter
for models 1--9.  The dotted lines show the two-component, exponential plus
Sersic-law fits to the surface density.  Again these are raw fits to the
luminous matter, ie.  no account is made of whether particles originally
belonged to the bulge, to the disk or to the satellite when performing the
decomposition of the profile.  Thus, we analyze the mass distribution in
much the same way as an observer would model the light distribution of a
face-on spiral galaxy. Points inside $r=0.15$ were excluded from the fit, as they are too
sensitive to the flattening of the initial bulge surface density profile.  The fits were made using a Levenberg-Marquardt nonlinear fitting algorithm to locate the $\chi^2$ minimum.  All model parameters were allowed to change. The
entire set of fitted parameters and the bulge-to-disk mass ratio derived
from the fits are tabulated in Table~\ref{Tab:FitResults}. 
The final $n$ increases with the
satellite mass, being slightly higher in the retrograde cases (7\%) than in
the prograde cases.  $n$ reaches 4 for a satellite as massive as the
original bulge.
In order to compute the goddness of the fits we computed the reduced $\chi^2$ for fits in which $n$ was fixed at values ranging from 0.8 to 5.  
Figure~\ref{Fig:chi}a--j  shows the reduced $\chi^2$ vs $n$.  
arrows indicate the minimum $\chi^2$ values listed in Table~\ref{Tab:FitResults}.  The minimum values of $\chi^2$ together with the number of data points in 
the fit $n = 30$, imply a better than 99\% confidence that the errors are
not due to mismatch of the model to the data. Figure~\ref{Fig:chi}a--j demonstrates that the remnant bulges have indeed evolved away from the $n=1$ initial exponential profile, given that values of $n$ below the best-fit values are strongly excluded.  The $\chi^2$ profiles have well-defined minima, with the exception of the massive-satellite models H1D, H1R, K1R.  These models admit fits with $n \geq 4$.

\begin{figure*}[!htb]
    \begin{flushleft}
\psfig{figure=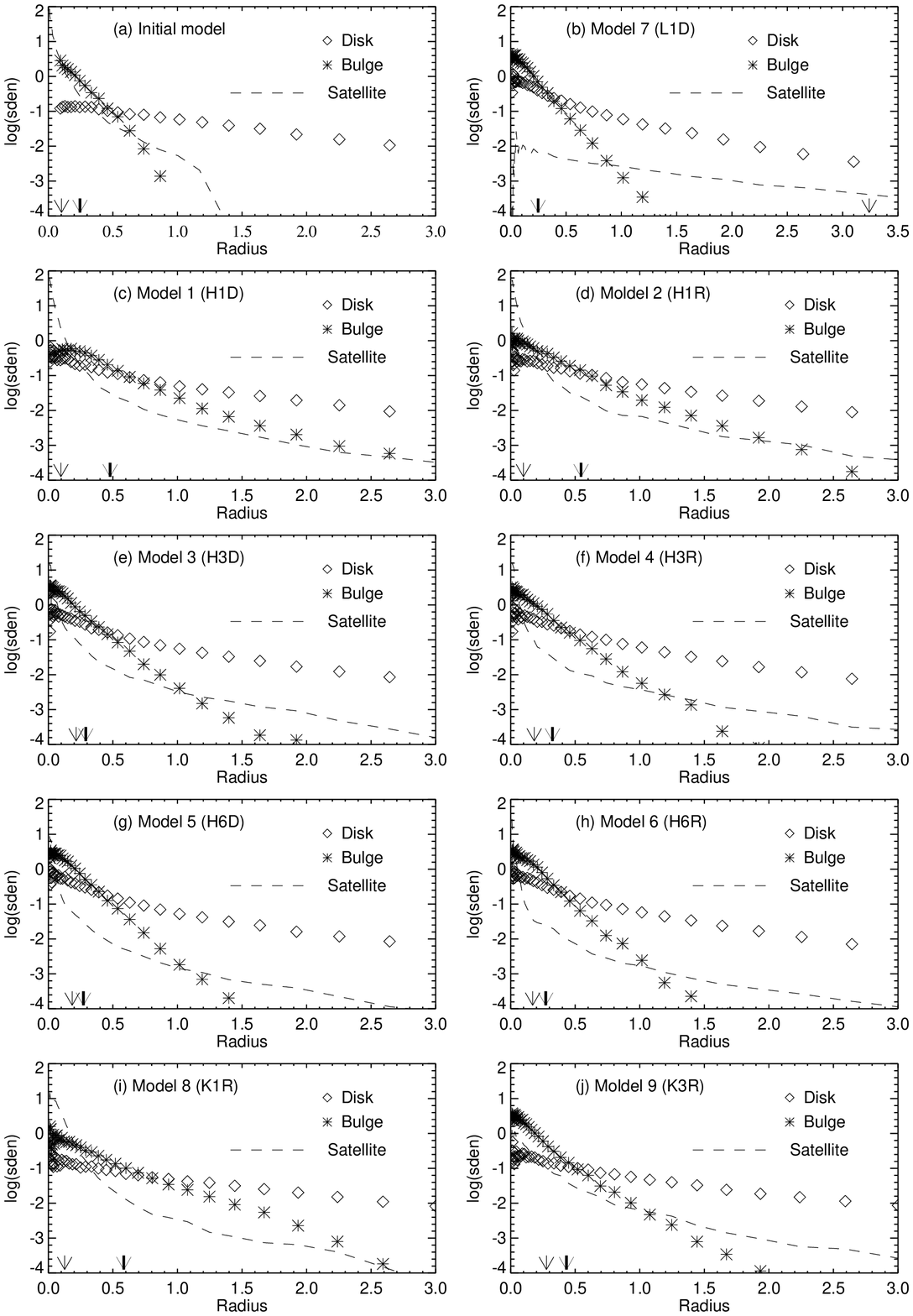,height=17.truecm,width=17.truecm}
              \caption[Fig:SBcomponents]{Radial surface density distributions of The various luminous components, after the merger.  {\it Asterisks: } bulge material.  {\it Diamonds: } disk material.  {\it Dotted line: } Satellite material.  Near the x-axis origin, {\it Thick arrow: } Half mass radius of the bulge matter distribution.  {\it Thin arrow: } Half mass radius of the satellite matter distribution.\label{Fig:SBcomponents}}
    \end{flushleft}
    \end{figure*}

The models displayed in figures~\ref{Fig:SBfits}c--h all correspond
to Hernquist satellites.  Models H1R and H3R were repeated using King model
satellites, as a basic test of the dependency of the result on the choice
of satellite model.  These are models 8 (K1R) and 9 (K3R) in
Table~\ref{Tab:ParsOrbit}.  Masses and half-mass radii for the satellites
of models K1R and K3R are identical to those of models H1R and H3R,
respectively.  The surface density profiles of models K1R and K3R
(Fig.~\ref{Fig:SBfits}i,j) are very similar to those of H1R, H3R, except 
at the very center.  The Sersic shape index $n$, given in each
figure, is nearly identical in models K1R and K3R as in models H1R and H3R,
respectively.  This suggests that the present results do not depend on a
fundamental way on the details of the internal structure of the satellite,
but mostly on the satellite total mass and mean density.  

\begin{figure*}[!htb]
    \begin{flushleft}
\psfig{figure=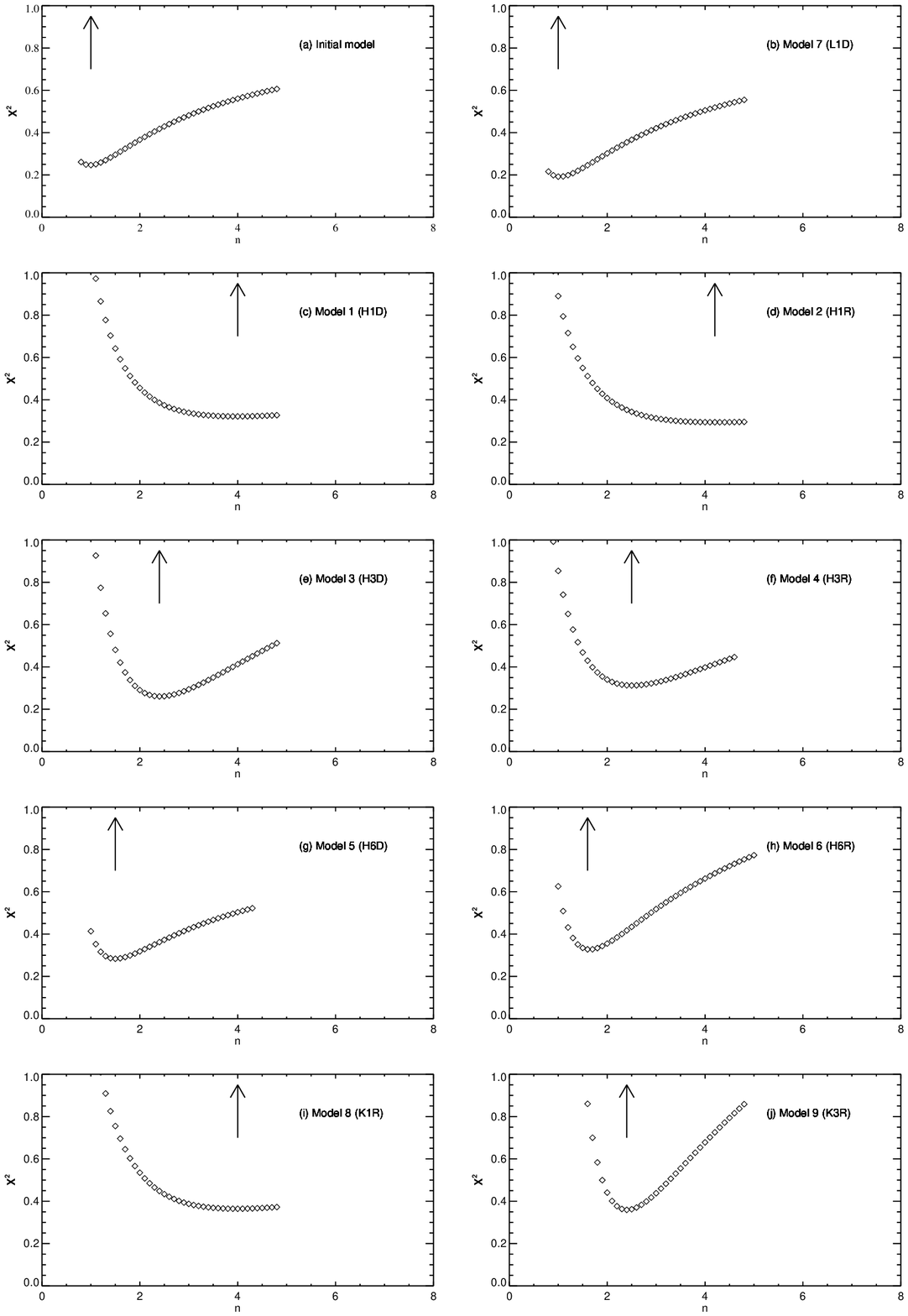,height=17.truecm,width=17.truecm}
              \caption[Fig:chi]{Reduced $\chi^{2}$ values for the differents fits. The mininum value of $\chi^{2}$ is indicated by arrows in each plot. \label{Fig:chi}}
    \end{flushleft}
    \end{figure*}

\begin{figure}[!htb]
    \begin{flushleft}
\psfig{figure=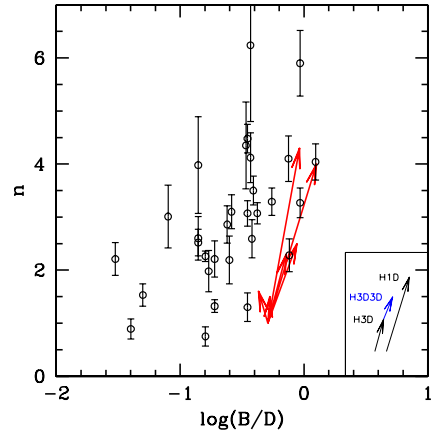,height=9.truecm,width=9.truecm}
              \caption[Fig:NvsMrat]{Growth vectors in the $n$--$\log(B/D)$ plane. Each arrow starts at the location of the original model and ends at the $n$ and $B/D$ derived from the two-component fit to the surface density profile after the merger. Growth vectors  for models H3D and H3D3D have been displaced to the left for clarity. They are displayed in arbitrary
units. {\it Points: } The distribution of $n$ vs. $\log(B/D)$ for bulges, from Fig.~5b of APB95.\label{Fig:NvsMrat} }
    \end{flushleft}
    \end{figure}

Figure~\ref{Fig:NvsMrat} shows growth vectors in the $n$--$\log(B/D)$
plane.  As discussed above, the bulge-to-disk ratios are those derived from
the Sersic plus exponential fits, i.e. they are the ones an observer would
derive.  Overall, bulges grow with the accretion process.  As satellite
accretion deposits mass onto the central bulge, it drives the increase of
the $n$ shape parameter of the surface density profile.  Plotted in the
same figure is the distribution of $n$ vs $B/D$ for the bulges in the BP94
sample, extracted from Figure~5b of APB95.  The match is encouraging, as
the growth vectors trace an increase of $n$ Sersic index with B/D just as
observed in the data; and most of the range of $n$ values displayed by the
data is obtained in the models.  Admitedly our models cover the high-mass
end of the $B/D$ space only.  Also, the growth vectors trace a somewhat
steeper slope than that of the distribution of data points.  Indeed,
accretion appears as being effective in driving bulges toward high values
of $n$.   Two conclusions can be drawn from this result. 
First, collisionless accretion of dense satellites onto disk-bulge-halo galaxies causes the bulge surface density profile to evolve toward higher-$n$ Sersic profiles.  That a merger of similar-mass systems leads to an $r^{1/4}$ profile is well known (Gerhard 1981, Barnes 1988).  The present mergers involve very unequal masses, thus our result is not a direct consequence of theirs, and indeed the final (total) mass distribution in our models differs substantially from the $r^{1/4}$ law.  The second conclusion is that bulges showing exponential surface brightness profiles may not have significantly grown via collisionless accretion of dense satellites.

As said in \S~\ref{Sec:Models}, we let the remnant of model H3D absorb a second satellite of the same mass as the first.  The inset to Figure~\ref{Fig:NvsMrat} shows the growth vector of the resulting model, H3D3D, together with those of H3D and H1D.  The second merger results in a similar increase in $n$ as the first merger, as well as a similar fractional growth of $B/D$.  The figure suggests that a subsequent merger with a similar satellite (total accreted mass in the three mergers equal to the initial bulge mass) would bring the final remnant to a $n$ and $B/D$ comparable to those of model H1D (single satellite with mass equal to the initial bulge mass).  This suggests that the evolution of bulges in the $n$--$B/D$ plane has little dependence on whether the mass is accreted in one event or peacemeal.  

The satellite needs to reach the galaxy center undisrupted for it to modify
the bulge shape parameter.  In the low density satellite case (model 7,
L1D), the satellite disrupted completely before reaching the bulge.  The
decomposition of the surface brightness profile in this case is shown in
Figure~\ref{Fig:SBfits}b.  The best fit for the bulge has $n=1$, the same
value as in the initial profile.  The mass distribution is however not
entirely undisturbed.  Comparison with the initial surface density profile
(Fig.~\ref{Fig:SBfits}a) shows that the final extrapolated disk
central surface density of model L1D is slightly lower, and, more
significantly, the central density of the bulge component is higher than in
the initial model.  This is surprising as the satellite has disrupted
completely before reaching the center.

We now analyze the contribution of each mass component to the final surface
density profiles.  Figure~\ref{Fig:SBcomponents} shows the surface density
distributions for matter originally belonging to bulge, disk and satellite,
for all of the merger models.  Panel $(a)$ gives the distributions in the
models before the merger.  Near the x-axis of each plot, vertical arrows
mark the half-mass radius $r_{1/2}$ for the distribution of bulge particles
(thick arrows) and that of satellite particles (thin arrows).

The distributions of particles initially belonging to the bulge and to the 
satellite develop  extended tails in all cases, with $r_{1/2}$ being 
larger for the more massive satellite cases.  For the bulge, the tail is a 
result of the absorption by dynamical friction of the orbital energy and 
angular momentum of the satellite, while for the satellite it is a result 
of stripping.   Matter from both the bulge and the satellite contribute to 
the final profile of the inner component deviating from the straight, 
exponential shape.  For the more massive satellite mergers, the central 
density of bulge material drops significantly, the center being filled 
with satellite material;  the central density of bulge particles is 
roughly unchanged for lower mass satellites.   

There is a third contribution to the final shape of the inner component 
deviating from an exponential shape -- that of the disk material.  In all 
cases, the surface density profile of particles originally belonging to 
the disk significantly deviates from an exponential law, curving up inward 
and showing a significantly higher central surface density than in the 
initial model.  Disk material has been dragged inward during the merger.  
When performing raw two-component fits such as those shown in  
Figure~\ref{Fig:SBfits}, matter originally belonging to the disk 
contributes to the inner component.  This occurs even in model L1D where the 
satellite has completely disrupted before reaching the center.  

How much damage does the initial bulge take as as result of the absorption
of orbital energy?  Figure~\ref{Fig:FracMasRad} gives $R_m$, the radius
enclosing a given fraction of the mass, for the bulge particles in the
initial and the final snapshots for models 1--7.  In models 1 (H1D) and 2
(H1R), the initial bulge expands at all radii, reflecting the effects of
the denser satellite on the relatively fragile bulge.  Smaller satellites
(models 3--7) deposit most of their energy in the outer layers.  In all
cases, the increase in $R_m$ traces the formation of an extended tail which
turns the initial exponential profile into an $n > 1$ Sersic profile.

\section{Internal kinematics of the merger remnants}
\label{Sec:Kinematics}

\begin{figure*}[!htb]
    \begin{flushleft}
\psfig{figure=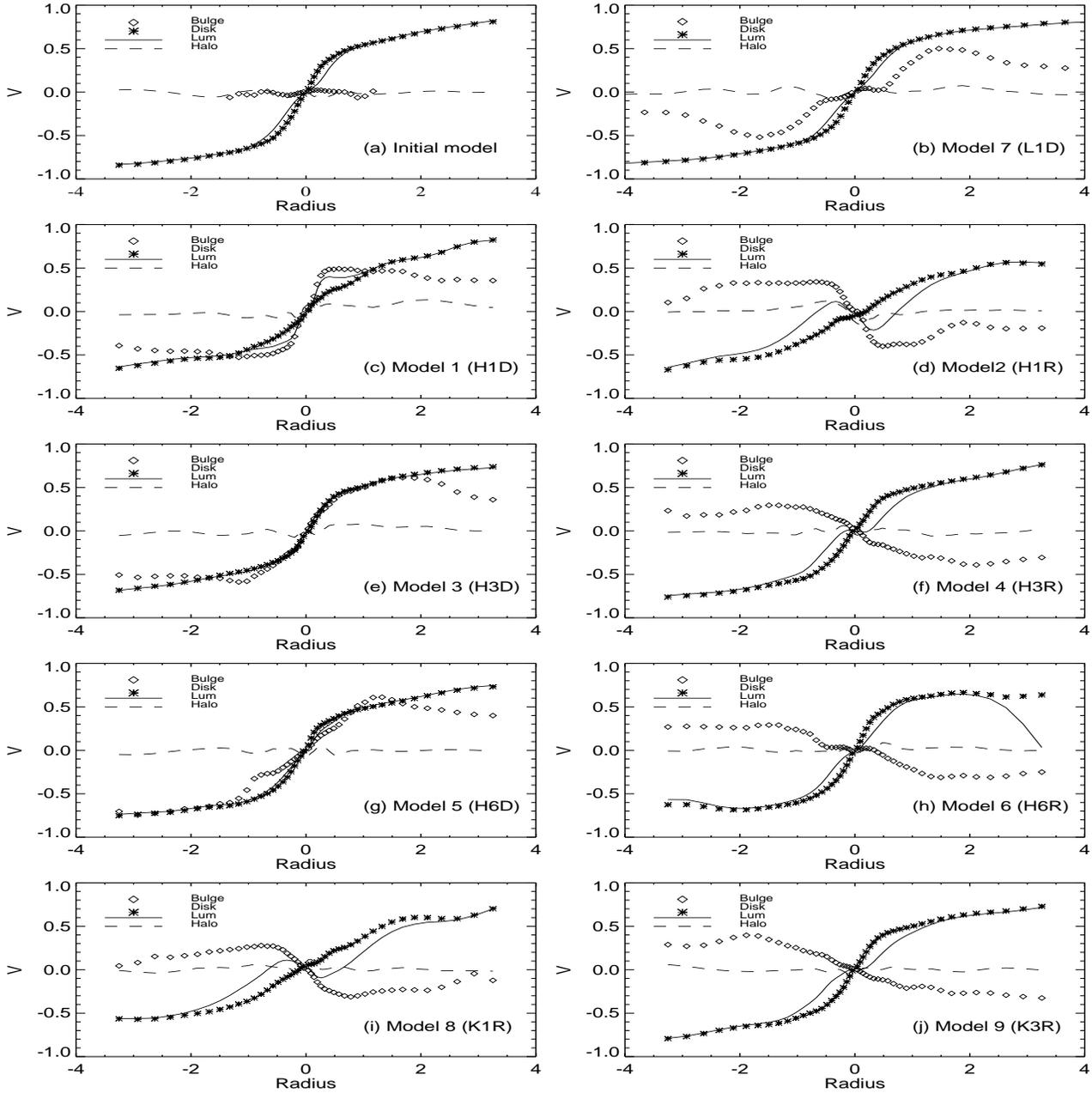,height=17.truecm,width=17.truecm}
              \caption[Fig:Rotcur]{Line-of-sight velocities of the models, observed from a viewing angle in the plane of the disk.  
{\it (a)}  Initial model. 
{\it (b)} Low-density model 7 after the merger is complete.  
{\it (c--h)} Models 1--6 after the merger is complete.  
{\it Stars: } disk material.  
{\it Dashed line: } Dark matter halo. 
{\it Solid line: } Luminous matter.\label{Fig:Rotcur} }
    \end{flushleft}
    \end{figure*}

\begin{figure*}[!htb]
    \begin{flushleft}
\psfig{figure=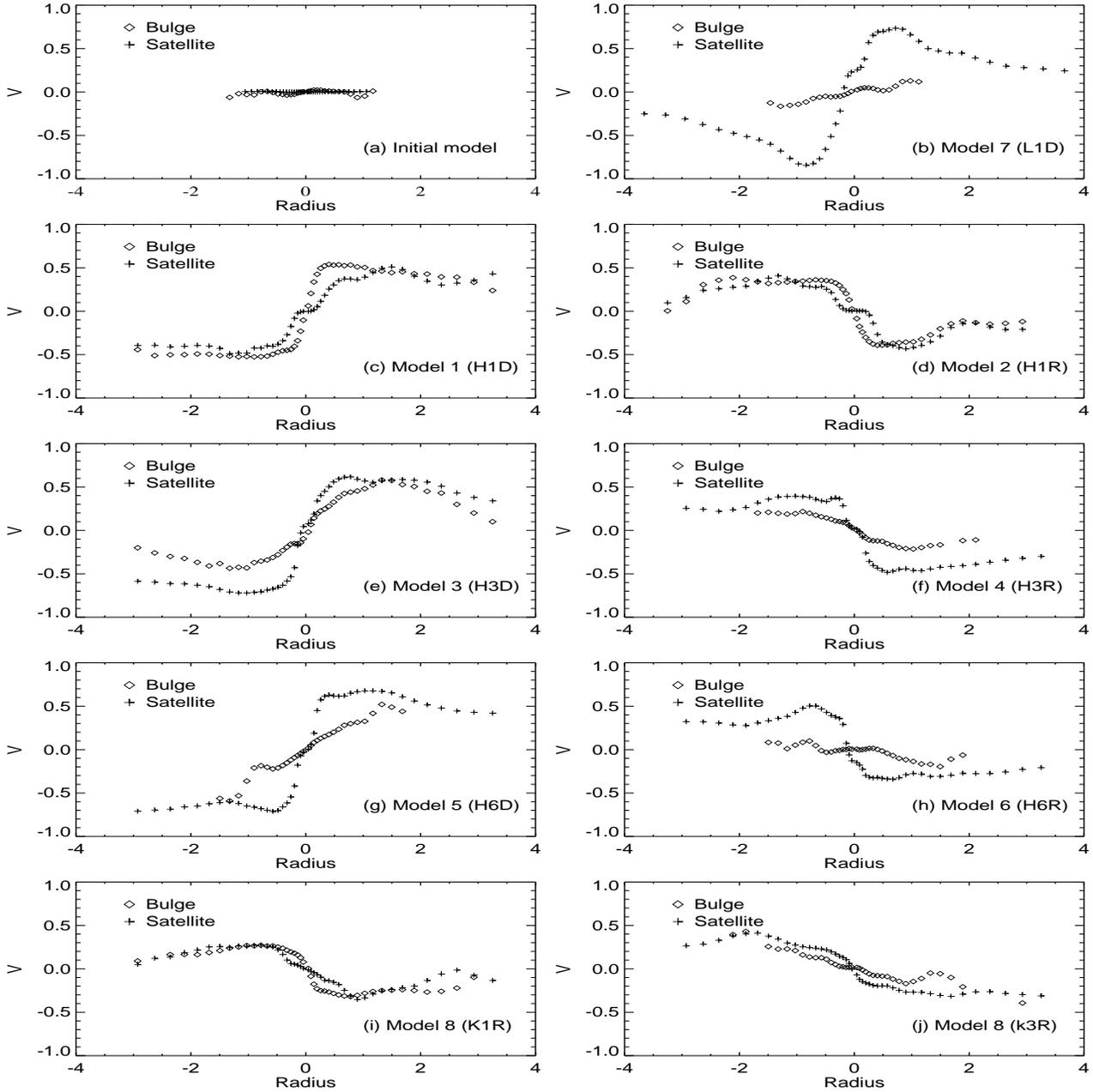,height=17.truecm,width=17.truecm}
              \caption[Fig:RotcurBulSat]{Line-of-sight velocities of the particles originally in the bulge (full line) and in the satellite (diamonts) after the merger is complete, from a viewing angle in the plane of the disk.\label{Fig:RotcurBulSat} }
    \end{flushleft}
    \end{figure*}

\begin{figure}[!htb]
    \begin{flushleft}
\psfig{figure=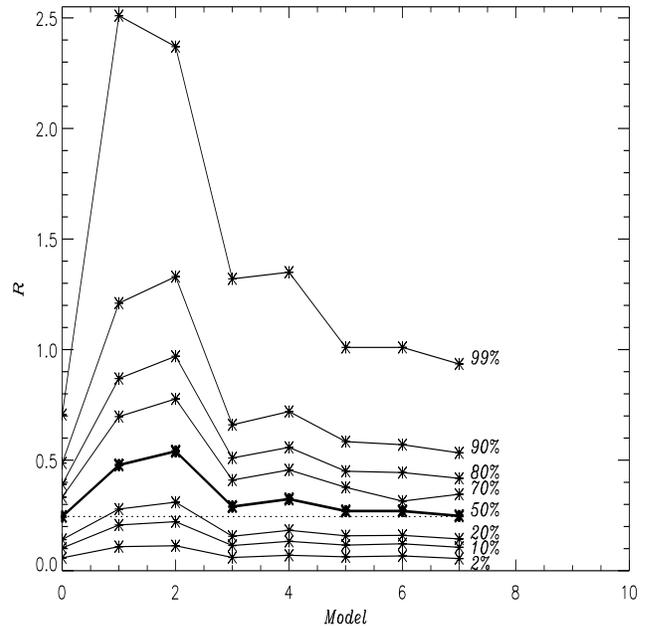,height=9.truecm,width=9.truecm}
              \caption[Fig:FracMasRad]{Radii enclosing a given \% of the 
mass for the distribution of particles initially belonging to the 
bulge.  The abscissa is the model number from Table~\ref{Tab:ParsOrbit}.
Model~0 represents the initial bulge set of $R_m$.  The horizontal 
dotted line is the radius initially enclosing 50\% of the bulge mass.\label{Fig:FracMasRad} }
    \end{flushleft}
    \end{figure}

The transformation brought forward by the merger modifies not only the 
galaxy's radial mass distribution but also its internal kinematics.  
Rotation curves of disk galaxies follow systematic patterns as a function 
of Hubble type, hence we may obtain additional diagnostics on the 
accretion by studying the rotation curves of the merger remnants.  The 
rotation curves of the initial model and of models 1--9 after the merger 
are shown in Figure~\ref{Fig:Rotcur}. These are line-of-sight velocities, 
measured along a virtual slit placed in the plane of the disk, with the 
disk seen edge-on.  Velocities for particles initially belonging to each 
of the galaxian components are shown with different symbols. For the 
post-merger rotation curves, bulge and satellite particles have been 
grouped together for clarity.  The individual rotation curves for bulge 
and satellite material are shown in Figure~\ref{Fig:RotcurBulSat}. 

We first analyze the direct models (Models 1, 3, 5, 7).  The overall
rotation curve of the luminous matter (Figure~\ref{Fig:Rotcur}, solid line)
is steeper in the model representing the earliest type (model H1D, largest
final bulge), becoming shallower for later types (model H6D).  This
behavior is similar to that observed in disk galaxies (eg.  Casertano \&
van Gorkom 1991).  The faster rotation of the model with the largest bulge
reflects not only the contribution of the massive bulge to the
gravitational potential but also the intrinsic fast rotation of the bulge. 
In the central parts this is higher than that of the disk material, whose
motion is partially pressure-supported due to heating: merger 1 has
produced an object that may resemble an S0 galaxy.  It is the material
originally belonging to the bulge that drives the fast inner rotation of
the final bulge (Fig.~\ref{Fig:RotcurBulSat}a), while the satellite
material rotates much slower.  The difference is due to the differences in
the density profile of bulge and satellite.  The bulge (King model) and
satellite (Hernquist model), despite having equal total masses, have
different central densities; when both merge, the satellite high-density
core eventually dominates the potential, driving bulge material to spin in
the satellite's wake.  This behavior is common to all mergers of galaxies
with unequal densities (Balcells \& Quinn 1990, Balcells \& Gonz\'alez
1998).  In models involving King satellites, the rotation curves of bulge
and satellite materials are more nearly similar
(Fig.~\ref{Fig:RotcurBulSat}i,j).  Smaller satellites have overall lower
effect on the total potential.  Stripped, high velocity material dominates
the rotation curve of the material originally belonging to the satellite
(Fig.~\ref{Fig:RotcurBulSat}g,h).

Retrograde models (nos. 2, 4, 6, 8, 9) show characteristic retrograde 
rotation of the bulge and satellite material. In the central region, the 
final bulge spins due to deposition of orbital angular momentum 
transported inward by the satellite, and the material belonging  to the 
initial bulge acquires rotation (our initial models featured non-rotating 
bulges). Outside the bulge (at roughly $r \ge 0.5$), stripped satellite 
particles and particles initially belonging to the bulge orbit as  test 
particles in the halo potential. This material may correspond to the 
retrograde moving groups in the Milky Way halo (Norris \& Ryan 1989, 
Majewski 1992), but would be unobservable in unresolved galaxies due to 
its low surface brightness.  

The counterrotation of the final bulge is common to all our retrograde 
mergers.  If the initial bulges had had direct rotation, we expect that 
the final bulge  would not counterrotate in the low-mass mergers (models 
H3R, H6R, K3R, Fig.~\ref{Fig:Rotcur}f,h,j), but new models are needed to verify 
this.  For the low-mass retrograde mergers, the fast counterrotating 
velocity of the satellite material (Fig.~\ref{Fig:RotcurBulSat}h), 
can be detected with line-profile spectral analysis techniques such as 
unresolved Gaussian decomposition (Kuijken \& Merrifield 1993).  
These simulations show that the 
counterrotation found in the bulge of NGC 7331 (Prada et al. 1996) could 
correspond to the remnant of an accreted low-mass satellite.  Bulge 
counterrotation has also been found in NGC 2841 (Silchenko et al 1997), 
but is otherwise uncommon in spiral and S0 galaxies (Kuijken et al. 1996).  

\section{Effects on the disk}
\label{Sec:DiskHeating}

\begin{figure}[!htb]
    \begin{flushleft}
\psfig{figure=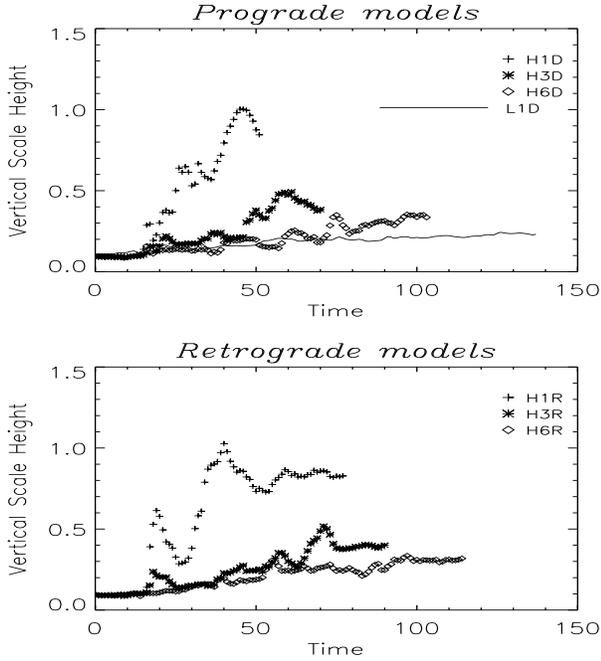,height=9.truecm,width=9.truecm}
              \caption[Fig:ZvsTime]{Time evolution of the disk scale height $z_{D}$ for the models. This scale length was measured at R=1 (see test for details).\label{Fig:ZvsTime} }
    \end{flushleft}
    \end{figure}

It is well known that mergers with dense satellites heat up the disk 
(QHF93), and the effect is clearly seen in Figures \ref{Fig:Snap1} and 
\ref{Fig:Snap5}.  Figure~\ref{Fig:ZvsTime} shows the disk scale height
$z_D$ for all models as a function of time.  $z_D$ has been measured at the
galactocentric distance of $R=1$ which corresponds to one initial disk
scale length.  The thickness of the disk increases with the mass of the
satellite galaxy.  Satellites with the same mass as that of the bulge
increase $z_D$ by about 2.5 times more than satellites with $1/3$ of the
bulge mass.  No significant differences can be found between prograde an
retrograde models.  Figure~\ref{Fig:ZvsTime} gives also the thickness of
the disk for the low-density satellite model (L1R).  The increment of
$z_D$ in model L1R is about 3.4 times less than the increment presented by
models H1D and H1R in which the satellite has the same mass but higher density. 
Our thickening results are comparable to those of QHF93.

A second important effect of the merger is to increase the disk scale
length.  $h$ increases by 10\% for the smaller, increasing to 60\% for the
largest satellites (Table~\ref{Tab:FitResults}).  The shallower disk
profile is a result of outward transport of disk material in the outer
parts, combined with inward transport to the bulge in the inner parts. 
Because the bulge effective radius does not increase strongly during the
merger, the ratio $r_e/h$ decreases as a result of the merger.  The final
$r_e/h$ does not scale with the final $B/D$ in any systematic way.  We
discuss the significance of these results in \S~\ref{Sec:Discussion}.

\section{Discussion}
\label{Sec:Discussion}

Our simulations show that, as a spiral galaxy grows its bulge by accretion 
of small satellites,  the bulge surface brightness profile quickly evolves 
from an $n=1$, exponential profile, to a  profile approaching $n=4$.  
Growth vectors on the $n$ vs $B/D$ plane suggest that the dependency of 
$n$ on Hubble type found by APB95 could be the result of satellite 
accretion.  

 The result goes beyond being just a manifestation of the known evolution of violently relaxing systems toward the $r^{1/4}$ law (eg. van Albada 1982).  Indeed, gravitational matter in our systems includes disk and halo in addition to the bulge.  Taken as a whole, none of the final remnants approach the $r^{1/4}$ law, especially so the luminous components.  Our results demonstrate that, only that specific subsystem identified with the central bulge,  gradually evolves toward $r^{1/4}$, while the rest of the luminous matter keeps its exponential surface density distribution.

Collisionless processes alone are involved in the transformation. 
Satellites that do not entirely disrupt during the merger are needed for
the process to operate with the efficiency shown here.  If the satellite
disrupts the bulge surface brightness profile remains undisturbed. 
However, the transformation does not rely on the satellite having an
$r^{1/4}$ to begin with: the evolution of the surface brightness profile is
similar whether the satellite is modeled with a Hernquist profile or a King
profile, suggesting that, as long as the satellite does not disrupt during
the merger, accretion-driven bulge growth makes the bulge evolve from $n=1$
toward $n=4$ profile shapes.  Indeed, the evolution of the bulge surface
brightness profiles is driven more by the puffing up of the bulge material
by the absorption of orbital energy and angular momentum of the satellite
than by the deposition of the satellite's high-density cusp in the remnant
center.  We expect that the details of the central surface brightness
profiles do depend on the shape of the central density profile of the
satellite (eg.  compare the central profiles of H1R to K1R, and H3R to K3R,
Fig.~\ref{Fig:SBfits}), but the effects on $n$ are small.  We do not make
quantitative predictions on the resulting central densities because our
King models deviate from exponentials within $r=0.16$, and because
softening limits the ability of the models to accurately reproduce central
densities.

\begin{table*}
\begin{center}
\caption{ \label{Tab:ParsOrbit} Orbital parameters for the merger 
experiments}
\begin{tabular}{ccccccc}
\hline
\hline
Model & Code & $M_{Sat}/M_B$ & $r_{1/2}^{Sat}$ & $V_{r}$ & $V_{\theta}$& 
Inclin (deg)\\
(1)   & (2)  & (3)           & (4)             & (5)     & (6)         & 
(7)         \\
\hline
1 & H1D & 1     & 0.243 & -0.001 & 0.222 & 30  \\
2 & H1R & 1     & 0.243 & -0.001 & 0.222 & 150 \\
3 & H3D & 0.333 & 0.102 & -0.025 & 0.227 & 30  \\
4 & H3R & 0.333 & 0.102 & -0.025 & 0.227 & 150 \\
5 & H6D & 0.167 & 0.068 & -0.115 & 0.219 & 30  \\
6 & H6R & 0.167 & 0.068 & -0.115 & 0.219 & 150 \\
7 & L1D & 1     & 0.347 & -0.011 & 0.222 & 30  \\
8 & K1R & 1     & 0.243 & -0.001 & 0.222 & 150 \\
9 & K3R & 0.333 & 0.102 & -0.025 & 0.227 & 150 \\
\hline
\end{tabular}
\end{center}
\tiny
Description of the columns: (1) Model number.  (2) Model
code.  (3) Initial mass ratio between satellite and bulge.  (4) Initial
half-mass radius of the satellite.  (5) and (6) Radial and tangential
velocity components of the relative orbit.  (7) Initial angle between the
orbital angular momentum and the disk spin.
\end{table*}

Mixing of the initial populations occurs not only due to heating but also
because the radial redistribution of disk material toward the center makes
some material initially belonging to the disk to be interpreted by the
observer as belonging to the bulge when fitting the disk with a standard
exponential profile.  This mixing may contribute to the extreme color
similarity between bulges and inner disks (Terndrup et al. 1994, Peletier
\& Balcells 1996).  Mixing does not occur at the very center, where the
satellite core ends up, hence on purely stellar dynamical grounds we should
expect a color signature there.  Detailed interpretation however is made
uncertain due to the presence of dust and star forming processes.  The
color structure in the inner few 100 pc of bulges is now known to be far
from simple, with steep reddening profiles and hidden star formation
processes revealed by HST/NICMOS-WFPC2 imaging (Peletier et al. 1999). 
Gas in the disk, not modeled in our simulations, is likely to respond more
violently to the accretion and pile up at the center (Barnes \& Hernquist
1996) probably triggering star formation and chemical enrichment.

The results are somewhat affected by the choices made for the simulation,
such as the bulge rotation, the satellite rotation or the types of merger
orbit.  Rotation in the initial bulge, not included in our models, would
probably make the prograde and the retrograde cases more different from
each other, broadening the range of evolution vectors in
Figure~\ref{Fig:NvsMrat}.  Otherwise, rotation is not likely to
significantly affect the evolution of $n$.  Differences may occur if the
bulge, or the satellite, had a central black hole.  Massive black holes are
now virtually confirmed in a number of galaxies, and they might be
by-products of the initial formation of the bulge (Rees 1993).  Central
black holes would probably affect both the final density profile and the
rotation of remnant.  Modeling disk galaxy mergers with smaller initial
bulges would be interesting, although difficult computationally.

\begin{table*}
\begin{center}
\caption{ \label{Tab:FitResults} Fitted parameters for Sersic plus 
exponential "bulge-disk" decomposition of final models} 
\begin{tabular}{cccccccccc}
\hline
\hline
Model & Code & \multicolumn {3}{c} {Bulge} & & \multicolumn {2}{c} {Disk} 
                                        & & B/D \\
\cline{3-5}\cline{7-8}
 && $log(I_{e})$ & $r_{e}$ & $n$ &  & $log(I_{o})$ & $h$ &  &  \\
(1)&(2)&(3) & (4) & (5) & & (6) & (7) & & (8) \\
\hline
Initial& ... &  0.04$\pm$0.05 & 0.21$\pm$0.02 & 1.05$\pm$0.13& & -0.80$\pm$ 0.03& 1.0$\pm$0.03& & 0.51  \\
1      & H1D & -0.05$\pm$0.01 & 0.23$\pm$0.01 & 4.03$\pm$0.11& & -1.23$\pm$ 0.02& 1.63$\pm$0.08& & 1.26 \\
2      & H1R & -0.21$\pm$0.06 & 0.23$\pm$0.01 & 4.30$\pm$0.23& & -1.03$\pm$ 0.04& 1.20$\pm$0.07& & 0.92 \\
3      & H3D & -0.20$\pm$0.03 & 0.24$\pm$0.05 & 2.35$\pm$0.15& & -1.12$\pm$ 0.05& 1.45$\pm$0.05& & 0.72 \\
4      & H3R & -0.18$\pm$0.04 & 0.22$\pm$0.03 & 2.51$\pm$0.09& & -0.90$\pm$ 0.03& 1.11$\pm$0.06& & 0.85 \\
5      & H6D & -0.12$\pm$0.02 & 0.21$\pm$0.01 & 1.53$\pm$0.12& & -0.96$\pm$ 0.01& 1.19$\pm$0.03& & 0.53 \\
6      & H6R & -0.18$\pm$0.03 & 0.20$\pm$0.02 & 1.65$\pm$0.16& & -0.92$\pm$ 0.06& 1.13$\pm$0.04& & 0.43 \\
7      & L1D & -0.01$\pm$0.01 & 0.21$\pm$0.02 & 1.03$\pm$0.11& & -0.89$\pm$ 0.07& 1.12$\pm$0.03& & 0.51 \\
8      & K1R & -0.05$\pm$0.02 & 0.23$\pm$0.02 & 4.01$\pm$0.13& & -0.90$\pm$ 0.03& 1.05$\pm$0.02& & 1.24 \\
9      & K3R & -0.10$\pm$0.03 & 0.23$\pm$0.04 & 2.53$\pm$0.17& & -0.90$\pm$ 0.02& 1.06$\pm$0.03& & 0.88 \\
\hline
\end{tabular}
\end{center}
\tiny
Description of the columns: (3) Bulge effective surface density, (4) Bulge effective radius, 
(5) Bulge profile index, (6) Disk central intensity, (7) Disk scale length,
(8) Bulge-to-disk final mass ratio as derived from the Sersic plus
exponential double fit.
\end{table*}

The lack of bulge initial rotation and the higher density of the satellite 
both contribute to making counterrotation more prevalent in the models than 
they are likely to be in reality, and it is not a prediction of our models 
that half of all bulges should contain counterrotating components.  
Quantifying merger-induced bulge counterrotation will require a 
comprehensive set of simulations that include a realistic distribution of 
satellite orbit orientations (Zaritsky et al. 1997).  

The accretion of dense satellites is less effective in making the bulge's
effective radius $r_e$ grow than it is at making the disk scale length
$h_D$ grow (Table~\ref{Tab:FitResults} from absorption of orbital energy
and angular momentum.  The decrease of $r_e / h_D$ with each merger is at
odds with the observed increase of $r_e / h_D$ toward early types in real
galaxies (Graham \& Prieto 1999).  Two explanations for this mismatch are
possible.  First we note that the dense satellites deposit most of their
mass in the center of the bulge, while the low density model (LD1) deposits
all of its mass outside the bulge.  We may envision that for intermediate
density satellites the mass is deposited throughout the bulge driving a
stronger increase in the bulge's $r_e$.  Second, we discuss below that the 
surviving disk in in fact a thick disk.  De Grijs \& Peletier (1997) 
measure the scale lengths of thick and thin disks in edge-on galaxies, 
showing that $h_{thick} >> h_{thin}$.  This may explain that $r_e / h_D$ 
decreases in our models, where $h_D = H_{thick}$, while it increases in 
real galaxies, where $h_D = h_{thin}$.  

The disks thicken during the process.   The exact values shown in Fig. 10 are sensitive to the choice of energy and inclination of the initial orbits, with higher thickening expected of more inclined, and more energetic, orbits. 
Discussion of the disk vertical heating effects are beyond the scope of
this paper, and are extensively addressed in eg.  QHF93.  Our
thickening results are similar to theirs, and may be interpreted as
indicating that the merger contributes to the growth of a thick disk.
Lowering the satellite mass to reduce the impact on the disk helps a
bit.  Our low-mass satellites impart less damage to the disk.  But
their effect on the surface density of the bulge is also limited ($n=1$
to $n \sim 1.5$).  It is plausible, but not obvious a priori, that a
sequence of such mergers may drive $n$ to keep growing.  Lowering the
satellite mass even further is of limited use, as dynamical friction
becomes inefficient.  Satellites with mass 0.1 times the bulge mass
have merger times approaching one Hubble time.  We ran a multi-merger
case in which 10 satellites with mass 0.1 times the bulge mass evolved
around the disk galaxy.  Weak dynamical friction and
satellite-satellite interactions prevented the merger over one Hubble
time.

Therefore, disk thickening is probably unavoidable, and, because we start
out with an already formed disk, the present scenario of bulge growth is
variation on the "bulge before the disk" theme.  Indeed, for such galaxy to
match present day disk galaxies, a thin disk must be rebuilt out of the gas
remaining after the last merger.  The formation of the thick disk is itself 
not a problem for this scenario, as thick disks are present in early type 
disk galaxies and absent in late-types (de Grijs \& Peletier 1997).  

Merger-driven bulge formation prior to the disk (e.g. Kauffmann et al. 
1994) is at odds with the fact that Sc and later-type spirals have
exponential bulges: our experiments indicate that exponential bulges are
extremely fragile against accretion and merging.  The seeds of the central
bulges of spiral galaxies must be given by events unrelated to accretion of
smaller galaxies.

The efficiency of the increase of $n$ (Fig.~\ref{Fig:NvsMrat}) places 
limits on the fractional growth of $n=1-2$  bulges by collisionless 
accretion of dense satellites.  Our results suggest that collisionless 
accretion of dense satellites bear the signature of a non-exponential 
bulge surface density profile.  Such signature lends itself to tests 
requiring imaging data only, hence can be be used to constrain the merger 
history of galaxies at high cosmological distances.  

\section{Summary}
\label{Sec:Summary}

The collisionless accretion of dense satellites onto disk galaxies drives
the growth of the bulge and an increase in the $n$ index of the Sersic fit
to the bulge surface density profile.  The mass of the bulge and $n$ grow
proportional to the satellite mass.  A single merger with a satellite as
massive as the bulge forms an $r^{1/4}$ bulge.  The range of $n$ values
obtained in the models matches that observed in bulges by APB95, which
points at accretion as a simple way of setting up the APB95
relation.  These results support the idea that bulges of late type spirals
which show exponential surface brightness profiles can evolve to bulges of
early type galaxies with $n>1$ by mergers with satellites galaxies.

We predict a fair amount of population mixing by expansion of the bulge 
material and piling up of disk material to the center.  

The rotation curves of the merger remnants are steeper for more massive 
bulges, in accordance with observations.  Retrograde mergers may lead to 
counterrotating bulges.

The models predict that thick disks form as by-products of the evolution
toward larger B/D, in accordance with the presence of thick disks in S0 and
Sa disk galaxies.  Matching to present day galaxies requires the rebuilding
of a thin disk out of remaining gas.

 The efficient transformation of exponential bulges by accretion suggests that 
collisionless mergers at high-$z$ are not responsible for the structure of bulges of late-type spirals.

\subsection{Acknowledgments}

JALA was supported by grant 20-56888.99 from the Schweizerischer Nationalfonds.

\end{document}